# Comparative Analysis of Non-Invasive and Invasive Tumor Treatment Fields: A Simulation Study

Minmin Wang[a,b]

[a] Key Laboratory of Biomedical Engineering of Education Ministry, Department of Biomedical Engineering, School of Biomedical Engineering and Instrument Science, Zhejiang University, Hangzhou, China

[b] Qiushi Academy for Advanced Studies, Zhejiang University, Hangzhou, China

**Abstract**

This study compares electric field and temperature distributions between non-invasive and invasive tumor treatment fields (TTFields). We employ four-layer spherical head models, representing the scalp, skull, cerebrospinal fluid, and brain, for simulation analysis. Non-invasive TTFields utilize scalp transducers, while invasive methods involve electrode implantation into tumors. Our findings underscore the advantages of invasive TTFields, showcasing their superior tumor-targeting abilities and reduced energy requirements. Furthermore, our analysis of brain tissue temperature changes in response to TTFields indicates that non-invasive TTFields primarily generate heat on the scalp, whereas implantation methods concentrate heat production within tumors, preserving normal brain tissue. In conclusion, invasive TTFields demonstrates potential for precise and effective tumor treatment. Its enhanced targeting capabilities and limited impact on healthy tissue make it a promising avenue for further research in the realm of cancer treatment.

Comparative Analysis of Non-Invasive and Invasive Tumor Treatment Fields: A Simulation Study

**1. Introduction**

Tumor Treatment Fields (TTFields) have emerged as a promising avenue in cancer treatment. This novel approach utilizes alternating electric fields of intermediate frequency (100-300 kHz) and low intensity (1-3 V/cm) to disrupt the proliferation of cancer cells [1]. Traditionally, TTFields employs two pairs of non-invasive transducer arrays positioned on the scalp to deliver electric fields. This non-invasive approach is generally considered more patient-friendly due to its lower risk profile [2]. However, non-invasive scalp-based TTFields, while promising, is not without its challenges. Undesirable side effects, such as temperature increases and scalp reactions, have been reported in some cases [3]. Thus, the treatment device (Optune®) monitors the temperature of the transducers and reduces the current whenever a transducer reaches 41 °C to mitigate the risk of overheating during TTFields [4,5]. Additionally, it is worth noting that a significant portion of the electric field generated during treatment may dissipate into non-essential regions [6,7].

In response to these challenges, invasive TTFields have emerged as a compelling alternative. This method entails the surgical implantation of electrodes either directly within or in close proximity to the tumor tissue, affording a more precise and localized delivery of TTFields [8]. Notably, invasive methods are distinguished by their capacity to achieve a focused electric fields within the tumor itself, potentially augmenting treatment efficacy [9,10]. However, it is imperative to acknowledge that this approach necessitates a surgical procedure, which may introduce inherent risks and complications associated with electrode implantation.

The selection between invasive and non-invasive TTFields modalities hinges upon a multitude of factors, encompassing tumor type, location, and the overall health status of the patient. Non-invasive methods are often preferred when targeting tumors in regions where surgery is challenging, while invasive approaches may be considered when a more targeted and intensified TTFields is desired [11,12]. This pivotal decision-making process commonly involves consultation with a multidisciplinary medical team, who consider the unique circumstances and treatment objectives of each individual patient. However, it is worth noting that there is a lack of study comparing the effectiveness of invasive and non-invasive methods through quantitative evaluation. This research gap emphasizes the need for further investigation to thoroughly assess the comparative benefits and effectiveness of these treatment approaches.

In this study, we leverage electric field simulations in tandem with temperature field simulations to quantitatively assess the dose-response relationships linked to both non-invasive and invasive TTFields. This combined analysis will provide a comprehensive perspective on the effectiveness and potential of these therapeutic modalities, contributing to the optimization and refinement of TTFields strategies in cancer treatment.

## 2. Materials and Methods

### 2.1 Construction of Spherical Head Model

As depicted in Figure 1A, the four-layer spherical head models were developed to represents the scalp, skull, CSF and brain. These layers corresponded to respective thicknesses of 5 mm, 7 mm, 2 mm, and 76 mm. A virtual lung tumor which comprises a tumor core and a surrounding tumor shell (outer radius of 10 mm and an inner core radius of 7 mm) was positioned in proximity to the parietal lobe of the brain. Geometries were defined and meshed by COMSOL Multiphysics. We employed suitable boundary conditions and input parameters to simulate the electric field distribution generated by different treatment modalities, including non-invasive and invasive approaches.

## 2.2 Non-Invasive and Invasive TTFields Setting

As illustrated in Figure 1A, we considered both invasive and non-invasive tumor settings in our simulations. For non-invasive TTFields settings, we configured the parameters to align with typical treatment conditions. This involved employing a pair of transducers positioned on opposite sides, with each pair featuring nine transducers, each having dimensions of 5 mm in diameter and 2 mm in thickness. These transducers were distinguished by a cumulative excitation current of 360 mA and operated at a frequency of 200 kHz.

In contrast, the invasive TTFields approach entailed the surgical implantation of electrodes directly into the tumor tissue. To assess the electric field distribution within the tumor, we simulated the presence of an implantable electrode, measuring 0.8 mm in width and featuring two stimulation points, each with a length of 6 mm. It is important to highlight that there were two variants of invasive methods. The first involved the return electrode through one of the stimulation points implanted within the tumor, with an excitation current of 30 mA (referred to as Invasive TTFields-I). The second variant entailed the return electrode being implanted beneath the scalp, with an application of 12.5 mA excitation current to each of the two stimulation points, resulting in a total excitation current of 25 mA (referred to as Invasive TTFields-II).

## 2.3 Simulation of Electric Field and Temperature Distribution

The induced electromagnetic wave lengths of TTFields in the modeled biological tissues are much larger than the size of the head. Thus Quasi-static Maxwell's equation is used in modeling. Consider human body as a homogeneous volume conductor $\Omega$, In the steady state, the electric potential distribution inside model $\Omega$ is governed by the complex quasi-static Laplace equation:

$$\nabla \cdot (\tilde{\sigma}\nabla V) = 0 \quad on\ \Omega \quad (1)$$

$$\tilde{\sigma} = \sigma + i\omega\varepsilon \quad (2)$$

Here, $\tilde{\sigma}$ and $V$ is the complex conductivity and the electrical potential in $\Omega$, respectively. $\sigma$ is the electrical conductivity, $\varepsilon$ is the permittivity.

The induced electric field $E$ was derived from the scalar potential as $E = -\nabla V$, and the current density $J$ was calculated from the electric field using Ohm's law as $J = \sigma E$. The physical properties of various components are detailed in Table 1 [13].

In conjunction with electric field simulations, we performed temperature field simulations to evaluate the thermal consequences of various tumor electric field treatment strategies. The Pennes bioheat equation along with relevant thermophysical properties were employed to model heat generation and distribution within the tumor and surrounding tissues:

$$\rho c \frac{\partial T}{\partial t} = \nabla \cdot (k\nabla T) + Q_{met} + Q_{blood} + Q_{JH} \quad (3)$$

$\rho$ denotes the density (kg/m3), $c$ signifies the specific heat [J/(kg °C)], $T$ denotes the temperature (°C), t signifies the time (s), and $k$ represents the thermal conductivity [W/(m °C)]. The initial term on the right-hand side encapsulates Fourier's law for thermal conduction, serving as the mathematical formulation for quantifying heat transfer through this mode of thermal propagation. The second term, designated as $Q_{met}$ (W/m$^3$), quantifies the heat generated due to metabolic activities within the tissue. For the purpose of our simulations, this term was held constant. The third term, $Q_{blood}$ (W/m$^3$), characterizes the energy exchanges between the tissue and the circulatory blood.

Furthermore, it is noteworthy that a uniform convection factor of 4 W/(m² °C) was assumed for all external boundaries, and an emissivity value of 1 was assigned to account for radiative heat exchange considerations.

A comprehensive comparative analysis was conducted to evaluate the effectiveness of non-invasive and invasive TTFields. Key outcome measures included electric field intensity distribution and temperature profiles were assessed across various treatment scenarios.

TABLE 1 The physical properties of each tissue and material

| Physical parameter | Scalp | Skull | CSF | Brain | Tumor shell | Tumor core | Gel | Transducers |
|---|---|---|---|---|---|---|---|---|
| Electric conductivity $\sigma$ (S/m) | 0.30 | 0.08 | 1.79 | 0.2 | 0.24 | 1.00 | 0.10 | 0 |
| Relative permittivity $\varepsilon_r$ (1) | 5000 | 200 | 110 | 3000 | 2000 | 110 | 100 | 10000 |
| Thermal conductivity k [W/(m°C)] | 0.34 | 1.16 | 0.60 | 0.565 | 0.550 | 0.550 | 0.60 | 6 |
| Specific heat c [J/(kg °C)] | 3150 | 1700 | 4200 | 3680 | 3600 | 3600 | 4186 | 527 |
| Density $\rho$ (kg/m³) | 1000 | 1500 | 1000 | 1036 | 1030 | 1030 | 1000 | 6060 |
| Blood perfusion rate $\omega^*(\times 10^{-3}\ 1/s)$ | 1.43 | 0.143 | 0 | 13.30 | 1.72 | 0 | NA | NA |
| Metabolic rate $Q_m$ (W/m³) | 363 | 70 | 0 | 16229 | 58000 | 0 | NA | NA |

## 3. Results

### 3.1 Electric Field Distribution of Non-Invasive and Invasive TTFields

As shown in figure 1B, when comparing the distribution of electric fields generated by various TTFields modalities within the brain, it becomes evident that when an average electric field distribution with an almost equivalent dose is achieved in the tumor shell (Non-Invasive TTFields: 108.76 V/m; Invasive TTFields-I: 102.72 V/m; Invasive TTFields-II: 111.94 V/m), the average electric field intensity produced by Non-Invasive TTFields in non-tumor regions significantly surpasses that of invasive methods (Non-Invasive TTFields: 77.77 V/m; Invasive TTFields-I: 0.91 V/m; Invasive TTFields-II: 2.61 V/m).

Furthermore, in comparison to the electric field distribution within tumor cores, the Invasive TTFields II method yields markedly lower electric fields than Invasive TTFields I (Invasive TTFields-I: 199.84 V/m; Invasive TTFields-II: 70.96 V/m).

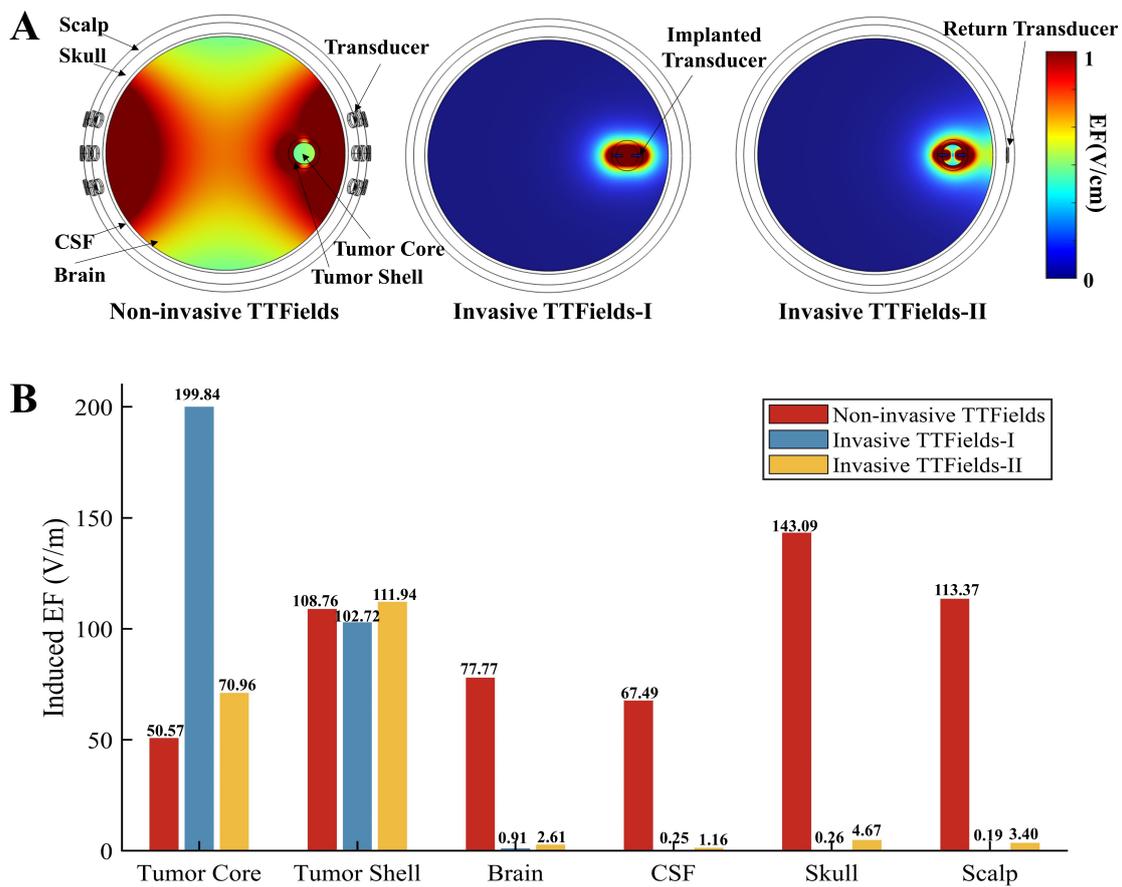

Figure 1 A.The distribution of induced electric fields within brain for Non-Invasive TTFields, Invasive TTFields-I and Invasive TTFields-II; B. The distribution of average induced electric fields within different tissue components

**3.2 Temperature Evaluation of Non-Invasive and Invasive TTFields**

In our analysis of temperature variations in human brain tissue subjected to various TTFields modalities, a discernible pattern emerges. As illustrated in figure 2, we observed that traditional non-invasive TTFields predominantly generate heat in the vicinity of the transducer on the scalp. Conversely, the implantation method primarily focuses heat production within the tumor area, exerting minimal impact on the adjacent normal brain tissue.

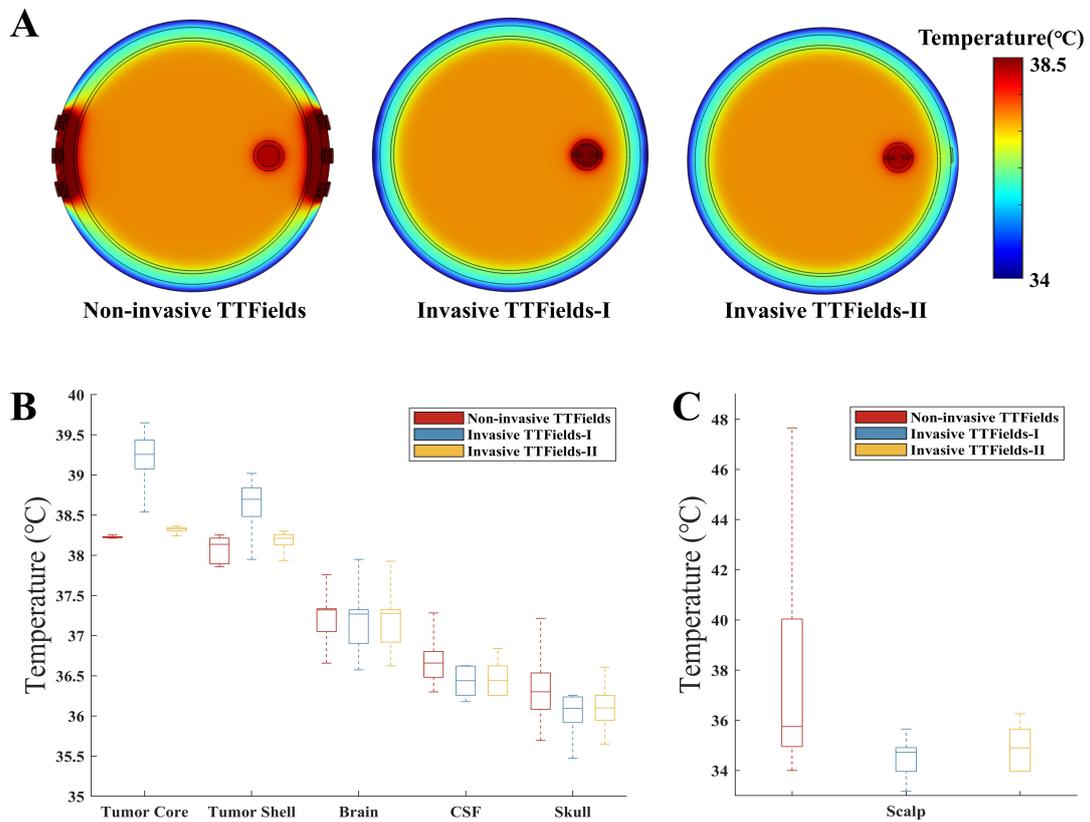

Figure 2. A. The distribution of maximum temperature within different tissue components for Non-Invasive TTFields, Invasive TTFields-I and Invasive TTFields-II; B-C. The distribution of temperature within different tissue components

## 4. Discussion

The comparative analysis conducted in this study has provided valuable insights into the electric field distribution and temperature profiles associated with non-invasive and invasive TTFields. These findings shed light on the crucial aspects of each treatment modality and their potential implications for optimizing tumor control strategies. The electric field distribution and temperature evaluation serve as vital parameters in assessing the efficacy and safety of these approaches.

Our results clearly indicate that invasive TTFields methods exhibit a superior ability to target the tumor site while requiring lower excitation energy compared to non-invasive methods. Moreover, the strategic positioning of return electrodes on the skull permits a heightened concentration of energy within the tumor shell. This observation underscores the potential feasibility of invasive TTFields therapy as an attractive option for precise and efficient tumor intervention. The enhanced targeting capability of invasive methods can result in more focused treatment, minimizing unintended effects on surrounding healthy tissue.

The Invasive TTF-II (return electrode was implanted under the scalp) emerges as a promising avenue for achieving selective and directional tumor electric field intervention. This hybrid approach harnesses the advantages of both modalities, enabling a synergistic electric field distribution within the tumor. The selective targeting achieved through this combination holds significant application prospects, potentially offering enhanced tumor control with minimized adverse effects. Further research and clinical investigations are warranted to explore the full potential of this hybrid approach and to refine its implementation.

It is essential to acknowledge the limitations of this study. The simulations and analyses conducted here are based on mathematical models and assumptions that may not fully capture the complexities of in vivo conditions. Clinical translation and validation of these findings are necessary to ascertain their real-world applicability and safety. Moreover, the study primarily focuses on brain tumor therapy, and the applicability of these findings to other tumor types requires further investigation.

Future research directions should involve experimental validation of the simulation results and the development of patient-specific treatment plans. Additionally, the investigation of novel electrode materials and configurations for invasive TTFields holds promise for optimizing targeting and reducing invasiveness. Furthermore, exploring adaptive treatment strategies that dynamically adjust electric field parameters based on real-time monitoring of tumor response could further enhance the effectiveness of TTFields.

## 5. Conclusion

In conclusion, this study underscores the significance of electric field distribution and temperature evaluation in evaluating TTFields. Invasive methods demonstrate enhanced targeting and lower excitation energy requirements, with Invasive TTFields-II showing potential for selective tumor electric field intervention. These findings contribute to the ongoing evolution of TTFields therapy and hold promise for improving the precision and efficacy of tumor treatment.

**Conflict of Interest**

All authors declare no competing interests.

**References**

1. Davies, A.M.; Weinberg, U.; Palti, Y. Tumor treating fields: a new frontier in cancer therapy. *Ann. N. Y. Acad. Sci.* 2013, *1291*, 86-95, doi:doi.org/10.1111/nyas.12112.
2. Rominiyi, O.; Vanderlinden, A.; Clenton, S.J.; Bridgewater, C.; Al-Tamimi, Y.; Collis, S.J. Tumour treating fields therapy for glioblastoma: current advances and future directions. *Br. J. Cancer* 2021, *124*, 697-709, doi:10.1038/s41416-020-01136-5.
3. Mehta, M.; Wen, P.; Nishikawa, R.; Reardon, D.; Peters, K. Critical review of the addition of tumor treating fields (TTFields) to the existing standard of care for newly diagnosed glioblastoma patients. *Crit. Rev. Oncol. Hematol.* 2017, *111*, 60-65, doi:doi.org/10.1016/j.critrevonc.2017.01.005.
4. Pohling, C.; Nguyen, H.; Chang, E.; Schubert, K.E.; Nie, Y.; Bashkirov, V.; Yamamoto, V.; Zeng, Y.; Stupp, R.; Schulte, R.W.; et al. Current status of the preclinical evaluation of alternating electric fields as a form of cancer therapy. *Bioelectrochemistry* 2023, *149*, 108287, doi:doi.org/10.1016/j.bioelechem.2022.108287.
5. Gentilal, N.; Miranda, P.C. Heat transfer during TTFields treatment: Influence of the uncertainty of the electric and thermal parameters on the predicted temperature distribution. *Comput. Methods Programs Biomed.* 2020, *196*, 105706, doi:doi.org/10.1016/j.cmpb.2020.105706.
6. Wenger, C.; Salvador, R.; Basser, P.J.; Miranda, P.C. The electric field distribution in the brain during TTFields therapy and its dependence on tissue dielectric properties and anatomy: a computational study. *Phys. Med. Biol.* 2015, *60*, 7339, doi:10.1088/0031-9155/60/18/7339.
7. Sun, Y.-S. Direct-Current Electric Field Distribution in the Brain for Tumor Treating Field Applications: A Simulation Study. *Comput. Math. Methods Med.* 2018, *2018*, 3829768, doi:10.1155/2018/3829768.
8. Di Sebastiano, A.R.; Deweyert, A.; Benoit, S.; Iredale, E.; Xu, H.; De Oliveira, C.; Wong, E.; Schmid, S.; Hebb, M.O. Preclinical outcomes of Intratumoral Modulation Therapy for glioblastoma. *Sci. Rep.* 2018, *8*, 7301, doi:10.1038/s41598-018-25639-7.
9. Deweyert, A.; Iredale, E.; Xu, H.; Wong, E.; Schmid, S.; Hebb, M.O. Diffuse intrinsic pontine glioma cells are vulnerable to low intensity electric fields delivered by intratumoral modulation therapy. *J. Neurooncol.* 2019, *143*, 49-56, doi:10.1007/s11060-019-03145-8.
10. Iredale, E.; Voigt, B.; Rankin, A.; Kim, K.W.; Chen, J.Z.; Schmid, S.; Hebb, M.O.; Peters, T.M.; Wong, E. Planning system for the optimization of electric field delivery using implanted electrodes for brain tumor control. *Med. Phys.* 2022, *49*, 6055-6067, doi:doi.org/10.1002/mp.15825.
11. Iredale, E.; Deweyert, A.; Hoover, D.A.; Chen, J.Z.; Schmid, S.; Hebb, M.O.; Peters, T.M.; Wong, E. Optimization of multi-electrode implant configurations and programming for the delivery of non-ablative electric fields in intratumoral modulation therapy. *Med. Phys.* 2020, *47*, 5441-5454, doi:doi.org/10.1002/mp.14496.

12. Jenkins, E.P.W.; Finch, A.; Gerigk, M.; Triantis, I.F.; Watts, C.; Malliaras, G.G. Electrotherapies for Glioblastoma. *Advanced Science* 2021, *8*, 2100978, doi:doi.org/10.1002/advs.202100978.
13. Gentilal, N.; Abend, E.; Naveh, A.; Marciano, T.; Balin, I.; Telepinsky, Y.; Miranda, P.C. Temperature and Impedance Variations During Tumor Treating Fields (TTFields) Treatment. *Front. Hum. Neurosci.* 2022, *16*, doi:10.3389/fnhum.2022.931818.